\documentclass[pra,aps,twocolumn,showpacs]{revtex4-2}
\usepackage{times,epsfig,amssymb,amsfonts,amsmath}
\newcommand{\ket}[1]{|#1\rangle}
\newcommand{\bra}[1]{\langle #1|}

\usepackage{graphicx}
\usepackage{subfigure}
\usepackage{upgreek}

\begin{document}
\title{Attaining near-ideal Dicke superradiance in expanded spatial domains}
	
	
\author{Jun Ren$^{1,2}$}
\email{renjun@hebtu.edu.cn}
\author{Shicheng Zhu$^2$}
\author{Z. D. Wang$^1$}
\email{zwang@hku.hk}

\affiliation{$^1$ Guangdong-Hong Kong Joint Laboratory of Quantum Matter, Department of Physics, and HK Institute of Quantum Science $\&$ Technology, The University of Hong Kong, Pokfulam Road, Hong Kong, People's Republic of China\\
$^2$ College of Physics and Hebei Key Laboratory of Photophysics Research and Application, Hebei Normal University, Shijiazhuang, Hebei 050024, People's Republic of China}

\begin{abstract}
		Dicke superradiance is essentially a case of correlated dissipation leading to the macroscopic quantum coherence. Superradiance for arrays of inverted emitters in free space requires interactions far beyond the nearest-neighbor, limiting its occurrence to small emitter-emitter distances. Epsilon-near-zero (ENZ) materials, which exhibit infinite effective wavelengths, can mediate long-range interactions between emitters. We investigate the superradiance properties of two ENZ structures, namely plasmonic waveguides and dielectric photonic crystals, and demonstrate their potential to support near-ideal Dicke superradiance across expanded spatial domains. We employ a general method that we have developed to assess the occurrence of superradiance, which is applicable to various coupling scenarios and only relies on the decoherence matrix. Furthermore, by numerically examining the emission dynamics of the few-emitter systems, we distinct the roles of quantum coherence at different stages of emission for the case of all-to-all interaction, and demonstrate that the maximum quantum coherence in the system can be determined using the maximum photon burst rate. The findings of this work have prospective applications in quantum information processing and light-matter interaction.

\end{abstract}
\maketitle

\section{Introduction} Dicke superradiance, originally described as the phenomenon that $N$ inverted emitters with identical interactions become phase-locked and emit photons in a short time with an intensity scaling as $N^2$ \cite{dicke54prl,eberly71pra,haroche82pr}, giving rise to emission burst. Superradiance has found extensive applications in diverse fields such as single-photon source \cite{duan01nat,chou04prl}, light harvesting \cite{mon97jpcb,scholes02cp,celardo14prb}, and quantum information processing \cite{duan01nat,reimann15prl,mh17prl,pog19prl}. Recent years have witnessed a lot of significant progress in exploring superradiance across various platforms such as ion traps \cite{dev96prl,esc01nat}, Rydberg gases \cite{haroche83prl,wang07pra,grimes17pra}, and superconducting systems \cite{van13sci,lamb16prb,wang20prl,orell22pra}. In the ideal scenario of Dicke superradiance, a large number of emitters are confined within a compact region, significantly smaller than the emission wavelength $\lambda$. This confinement ensures that the emitters become indistinguishable during the absorption or emission of a photon, resulting in the emergence of a Dicke state. Consequently, the Hilbert space, which would initially be $2^N$-dimensional, reduces to $N+1$ dimensions. Recent theoretical studies have expanded this superradiance theory to spatially extended and ordered emitter systems, such as atomic chains \cite{masson20prl,masson20prr,lopez23prl} as well as arrays \cite{rob21pra,ruks22prr,rubies22pra,masson22nc,sierra22prr}. These studies have demonstrated the significance of eigenvalues of an $N$-dimensional decoherence matrix. The elements of this matrix represent the incoherent emitter-emitter coupling (also known as dissipative coupling in certain literature), and play a crucial role in determining the minimal conditions for superradiance.

Most existing theories on superradiance focus on ordered atomic arrays in free space, where interactions between emitters decrease exponentially with the emitter-emitter distance. However, this limited perspective restricts the applicability of superradiance theory since non-uniform amplitudes and phases of the electromagnetic field surrounding emitters often lead to irregular emitter-emitter couplings and inhomogeneous single-emitter emission rates. Additionally, even for very short chains, superradiance bursts in free space can only be sustained when the emitter-emitter distance is sufficiently small \cite{masson20prl}. Ref.~\cite{masson22nc} establishes the minimum requirements for superradiance and provides a maximum inter-atomic separation that allows for the observation of superradiance in uniform atomic arrays. Moreover, a recent study \cite{mok23prl} demonstrates that superradiance cannot occur solely through nearest-neighbor atomic interactions. Consequently, the spatial domain at which superradiance manifests significantly shrinks in free space, and the intensity of the emission burst deviates significantly from the ideal Dicke superradiance.

Hence, it is desirable to develop a more comprehensive theory capable of assessing the existence of superradiance in complex scenarios, relying solely on the $N$-dimensional decoherence matrix. Additionally, two crucial requirements for achieving superradiance at large spatial domains are observed: long-range interactions and substantial dissipative coupling between distant emitters. Plasmonic or photonic waveguides have shown promise as facilitators of long-distance interactions and superradiance phenomena \cite{mar10plsnano,pus10prb,mlynek14nc,ren16prb,de16acsn,goban15prl,ren17jcp,ren19njp,li19prb,zhu23oe}. However, these systems often impose strict constraints on the spatial arrangement of emitters. Recent advancements in near-zero-index (NZI) materials, such as plasmonic epsilon-near-zero (ENZ) waveguides and all-dielectric ENZ photonic crystals, have overcome this limitation \cite{Ziolkowski04pre,alu07prb,Engheta13sci,maas13np,lib17np,Liedl23ar}. These materials, characterized by infinite wavelengths and vanishingly small index, enable all-to-all interactions and thus near-ideal realization of Dicke superradiance in extended spatial domains. Furthermore, as correlated dissipation can give rise to quantum coherence and superradiance, it is intriguing to investigate their roles in emission dynamics and elucidate the inherent connection between quantum resources and collective emissions.

In this work, we present a universal criterion for determining the occurrence of superradiance, which simplifies the analysis by focusing solely on the \textit{N}-dimensional decoherence matrix instead of computationally intensive calculations involving exponential growth with the number of emitters, \textit{N}. We apply this criterion to two ENZ systems, namely plasmonic waveguide and dielectric photonic crystal, both of which facilitate long-range emitter-emitter couplings. Our results demonstrate that these structures exhibit near-ideal Dicke superradiance over an expanded spatial domain. Furthermore, we perform numerical simulations to investigate the dynamics of systems with a few emitters under all-to-all coupling. By separately analyzing the emission dynamics with two distinct components: single-emitter and collective emission, along with the dynamics of quantum coherence, we have observed that quantum coherence can play completely different roles at different stages of emission, and the dynamics of quantum coherence and the collective emission are completely synchronized. Additionally, we establish an exact correspondence between the maximum quantum coherence and the maximum emission rate of superradiance, suggesting that the emission rate can serve as a potential measurement tool for quantum coherence.

The paper is organized as follows: Sec.~II presents the derivation of the minimal condition for superradiance, focusing on the \textit{N}-dimensional decoherence matrix, and investigates the relationship between the maximum emission rate and the second-order correlation function. In Sec.~III, we employ two ENZ materials, namely the plasmonic waveguide and the dielectric photonic crystal, to demonstrate the possibility of achieving near-ideal superradiance with multiple emitters in an expanded spatial domain. Section IV delves into the discussion of quantum coherence in the emission dynamics of all-inverted emitters under the all-to-all scenario. In Sec.~V we finally conclude.



\section{Theory of photon emission from multiple inverted emitters at early time}
In the study of the dynamic evolution of multi-emitter system in a weakly coupled environment, Born-Markov and rotating-wave approximations can be used. After tracing out the environment, the Lindblad master equation of emitter system can be written as \cite{dung02pra,gon11prl,ren19njp}
\begin{equation}
	\frac{\partial \rho}{\partial t}=\frac{i}{\hbar}[\rho, H]+\frac{1}{2} \sum_{i,j} \gamma_{ij}\left(2 \sigma_i \rho \sigma_j^{\dagger}-\rho \sigma_i^{\dagger} \sigma_j-\sigma_i^{\dagger} \sigma_j \rho\right),
\end{equation}
where $\sigma_i^{\dagger}$ and $\sigma_i$ are the raising and lowering operators of the $i$th emitter, respectively. The $\gamma_{ij}$ is the incoherent coupling between the $i$th and $j$th emitters, and all the $\gamma_{ij}$s together will be shown to control the emission process of multi-emitter system. The Hamiltonian of the emitter system is
\begin{equation}
	H=\hbar\omega_0 \sum_{i} \sigma_i^{\dagger} \sigma_i+\sum_{i \neq j} g_{ij} \sigma_i^{\dagger} \sigma_j,
\end{equation}
with $\omega_0$ being the transition frequency of the two-level emitters. The coherent coupling strength $g_{ij}$ has been shown to have a small impact on emission dynamic \cite{masson22nc}. In Eq.~(1), $\rho$ is the density matrix of the emitter system, $[\rho, H]$ and Lindblad term [the second term of the right-hand side of Eq.~(1)] govern the coherent and incoherent parts of the emitter dynamics, respectively.

Under the point dipole approximation, the coherent and incoherent interactions between the $i$th and $j$th emitters $g_{i j}$ and $\gamma_{ij}$ can be written as \cite{dung02pra,ren17jcp}
\begin{equation}
	g_{ij}=\frac{\omega _{0}^{2}}{\varepsilon _{0}\hbar {c^2}}\operatorname{Re}\left[ \vec{\mu }_{i}^{*}\cdot \overset{\leftrightarrow }{\mathop{G}}\,\left( {{{\vec{r}}}_i},{{{\vec{r}}}_j},\omega  \right)\cdot {{{\vec{\mu }}}_j} \right]
\end{equation}
and
\begin{equation}
	{{\gamma}_{ij}}=\frac{2\omega _{0}^{2}}{{{\varepsilon }_{0}}\hbar {{c}^{2}}}\operatorname{Im}\left[ \vec{\mu }_i^{*}\cdot \overset{\leftrightarrow }{\mathop{G}}\,\left( {{{\vec{r}}}_i},{{{\vec{r}}}_j},\omega  \right)\cdot {{{\vec{\mu }}}_j} \right],
\end{equation}
respectively. The $\vec{\mu}_i$ and $\vec{\mu}_j$ are the dipole moments of the $i$th and $j$th emitters, and the detailed calculation method of the electric Green's tensor $\overset{\leftrightarrow }{\mathop{G}}$ can be found in Sec.~III.

The total emission rate of the emitter system at time $t$ can be written as
\begin{equation}
	\gamma({t})=\gamma_{\rm s}(t)+\gamma_{\rm c}(t),
\end{equation}
where we define
\begin{equation}
\gamma_{\rm s}(t)\equiv\sum_{j}\gamma_{jj}\langle\hat{e}_{j}\rangle
\end{equation}
that governs the emission rate from single emitters, which originates from the diagonal elements of the density matrix $\rho(t)$ and decays exponentially, and 
\begin{equation}
	\gamma_{\rm c}(t)\equiv\sum_{i\neq j}\gamma_{i j}\langle\hat{\sigma}_{i}^{\dagger}\hat{\sigma}_{j}\rangle
\end{equation}
governs the collective emission, which causes the correlation-induced emission and originates from the non-diagonal elements of density matrix. The operators in Eqs.~(6) and (7) are $\hat{e}_{j}\equiv\ket{e_j}\bra{e_j}$, $\hat{\sigma}_{j}\equiv\ket{g_j}\bra{e_j}$, and $\hat{\sigma}_{j}^{\dagger}\equiv\ket{e_j}\bra{g_j}$, with $\ket{e_j}$ and $\ket{g_j}$ being the upper and lower energy levels of the $j$th emitter. Studying the single-emitter and collective emission dynamics independently can help us understand their roles in the overall emission process and how they relate to the superradiance and quantum resources in the system, as we can see in Sec.~IV.

Therefore, the superradiance occurs only when the increase rate of the total emission rate is larger than zero at the beginning of the emission, that is the first derivative ${\dot{\gamma}}(0)>0$. According to Refs.~\cite{rob21pra,sur21pra}, if the emitters are fully inverted initially, namely $\ket{\psi(0)}=\ket{e}^{\otimes N}$, the first derivative of $\langle\hat{e}_{j}\rangle$ and $\langle\hat{\sigma}_{i}^{\dagger}\hat{\sigma}_{j}\rangle$ at $t=0$ can be calculated as
\begin{equation}
	\frac{d\langle\hat{e}_{j}\rangle}{d t}(0)=-\gamma_{jj},\\
	\frac{d\langle{\hat{\sigma}_{i}^{\dagger}\hat{\sigma}_{j}}\rangle}{d t}(0)=\gamma_{ij}.
\end{equation}
Substituting Eq.~(8) into the first derivative of Eq.~(5), we have
\begin{equation}
	{\dot{\gamma}}(0)=-\sum_{j}\gamma_{jj}^2+\sum_{i\neq j}\gamma_{i j}^2.
\end{equation}
From Eqs.~(8) and (9) it can be inferred that the non-diagonal elements of the decoherence matrix $\boldsymbol \Gamma$, namely the dissipative couplings between emitters contribute positively to the superradiance whereas the diagonal elements (single-emitter emission rates) do the opposite. The competition between them determines whether or not the superradiance occurs. If we set the decoherence matrix $\boldsymbol \Gamma=\left( \gamma_{ij}\right) $ with the matrix elements $\gamma_{ij}$ being defined in Eq.~(4), the ${\dot{\gamma}}(0)$ can be written as a compact form
\begin{equation}
	{\dot{\gamma}}(0)={\rm Tr}\left(\boldsymbol \Gamma^2 \right) -2{\rm Tr}\left(\boldsymbol \Gamma_{\rm d}^2 \right),
\end{equation}
where ${\boldsymbol \Gamma}_{\rm d}\equiv{\rm diag(\gamma_{11},\gamma_{22},\cdots,\gamma_{NN})}$ is the diagonal matrix of $\boldsymbol \Gamma$. Consequently, the superradiance criterion $\dot{\gamma}(0)>0$ can be used to determine the minimal requirement for superradiance, which can be derived as 
\begin{equation}
	{\rm Tr}\left(\boldsymbol \Gamma^2 \right) > 2{\rm Tr}\left(\boldsymbol \Gamma_{\rm d}^2 \right).
\end{equation}

Here, it can be observed that superradiance in an all-inverted multi-emitter system can be determined without the need for dipole-dipole coupling [as shown in Eq.~(3)]. Therefore, we can state that whether superradiance occurs is independent with the coherent coupling between emitters. However, it should be pointed out that this does not imply that coherent coupling has no impact on the emission dynamics. Because the first derivatives of the expected values of the two operators shown in Eq.~(8) are decoupled only at the initial moment of emission, but are coupled with each other at all subsequent times, as shown in Refs.~\cite{rob21pra,sur21pra}.

Recent research \cite{masson20prl,masson20prr,masson22nc,lopez23prl} suggests that superradiance may be interpreted as the first emitting photon enhancing the rate of the second one, and this interpretation is based on the quantum jump operator profile \cite{kim00oc,car03pra}. In order to ascertain if superradiance occurs, they defined an easily computed second-order correlation function $g^{(2)}(0)$ of identical emitters at the beginning of the emission. Without assuming any special situations about the emitters' environment, we find that $g^{(2)}(0)$ can be calculated as
\begin{equation}
	g^{(2)}(0)=1+\frac{{\rm Tr}\left(\boldsymbol \Gamma^2 \right) -2{\rm Tr}\left(\boldsymbol \Gamma_{\rm d}^2 \right)}{\left({\rm Tr} \boldsymbol \Gamma\right) ^2},
\end{equation}
in which the meaning of $\boldsymbol \Gamma$ and $\boldsymbol \Gamma_{\rm d}$ are identical to those in Eq.~(10). The detailed procedure of jump operation method and the derivation of Eq.~(12) can be found in the Supplemental Material. Therefore, it is easy to find that the minimal superradiance condition of $g^{(2)}(0)>1$ is consistent with the condition indicated in Eq.~(11) we obtained. We employ $g^{(2)}(0)$ to identify the occurrence of the superradiance in the following study in this work because, as can be shown below, its value range is 0 to 2 [in contrast, ${\dot{\gamma}}(0)$ has a very large range of values and can go from negative to positive], and there is often an obvious positive connection between $g^{(2)}(0)$ and the emission rate.

Keep in mind that, other than assuming that all emitters are inverted initially and have the same transition frequency, we made no further assumptions while deriving the superradiance condition above. This result is important for those situations where emitters are in a complex environment rather than homogeneous free space, such as the plasmonic systems \cite{pus10prb,mar10plsnano,ren16prb}, where the decay and interaction are sensitive to the spatial arrangement of emitters, and these systems are frequently used to increase the emission rate of collective emitter systems. Therefore, as long as their decoherence matrix $\boldsymbol \Gamma$ is known, the results of Eqs.~(10)-(12) can be used to assess whether a multi-emitter system will undergo superradiance in general, and the dynamics of the whole $2^N$-dimensional system does not need to be calculated. Furthermore, as will be shown in Sec.~IV, $g^{(2)}(0)$ has an extensive connection with the emission dynamics, so it is also possible to calculate this value to assess the other emission characteristics, such as emission burst intensity, for some special cases and for a large number of emitters.

Next, we examine the initial second-order correlation function in several special cases. The first involves identical spontaneous decay but distinct emitter-emitter couplings. According to Cauchy-Schwarz inequality, we have
\begin{equation}
	\left(\sum_{i=1}^{N}\gamma_{ii}\right)^2 \leq N \left(\sum_{i=1}^{N}\gamma_{ii}^2\right),
\end{equation}
where the equality is taken if and only if all the diagonal elements $\gamma_{ii}$ are equal. Therefore, along with Eq.~(12) we can obtain a upper bound of $g^{(2)}(0)$
\begin{equation}
	g^{(2)}(0) \leq	1+\frac{\rm{Tr}(\boldsymbol \Gamma^2)}{(\rm{Tr}(\boldsymbol \Gamma))^2}-\frac{2}{N},
\end{equation}
where the equality is taken if and only if every emitter is in the same electromagnetic environment, resulting in an identical spontaneous decay rate. The right-hand side of Eq.~(14) is the same as in the existing results \cite{masson22nc,lopez23prl}, where the free-space environment is assumed. As indicated in Ref.~\cite{masson22nc}, for a constant emitter number $N$, the $g^{(2)}(0)$ is proportional to the variance of the eigenvalues of matrix $\boldsymbol \Gamma$ when the single emitter emission rate is identical. However, when the emitters' decay rates diverge, as illustrated in the Supplemental Material, this criterion is violated.

The second special case is the identical spontaneous decay and emitter-emitter coupling. We refer to this scenario as an all-to-all interaction since any two emitters will interact in the same way. We now assume that the emitters' spontaneous decay rates are equal since the amplitude of electric field in the emitter's region is uniform. Furthermore, we also assume that the phase of electric field keeps invariant in space and thus the dissipative couplings are also identical, but less than 1. If we set the coupling between any two emitters as $\alpha$ ($\alpha<1$), the normalized dissipative matrix has the form of
\begin{equation}
	\begin{split}
		&\boldsymbol \Gamma=
		\begin{pmatrix} 1 & \alpha & \cdots & \alpha  \\ \alpha & 1 & \cdots & \alpha \\ \vdots & \vdots & \ddots & \vdots \\ \alpha & \alpha & \cdots & 1 \end{pmatrix},
	\end{split}
\end{equation}
and according to Eq.~(12) the second-order function at the initial time $g^{(2)}(0)$ can be calculated as
\begin{equation}
	g^{(2)}(0)
	=\frac{N-1}{N}(1+\alpha^2).
\end{equation}
Obviously, to satisfy the superradiance criterion $g^{(2)}(0)>1$ in this scenario, $\alpha$ needs to be greater than $1/\sqrt{N-1}$ for a given $N$. Of course, this condition is readily met when $N$ is very large, and the actual challenge lies in achieving the all-to-all interaction in practical scenarios. In Sec.~III, we will go through in detail how to create this almost perfect all-to-all interaction situation.

\begin{figure}[t]
	\epsfig{figure=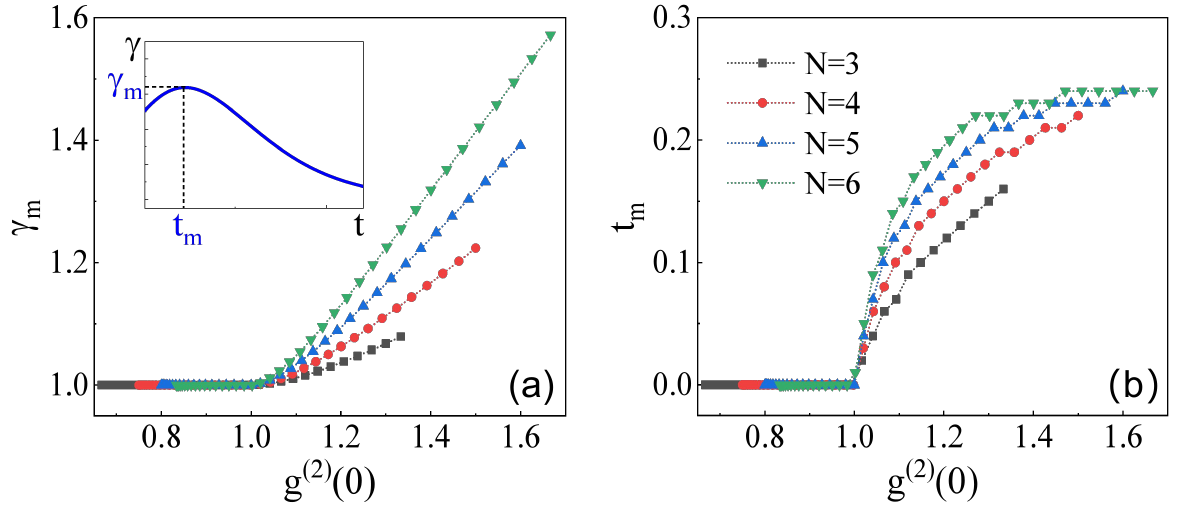,width=0.45\textwidth}
	\caption{(Color online) Relationship between maximum emission rate $\gamma_{\rm m}$ (a), its corresponding time $t_{\rm m}$ (b) with the second-order correlation $g^{(2)}(0)$ under different emitter number $N$. Inset: A schematic diagram of the relationship between the maximal emission rate of superradiance and its corresponding time. }
\end{figure}

When superradiance occurs, there are two important characteristic indices for the emission dynamics, the maximum total emission rate $\gamma_{\rm m}$, and its corresponding time $t_{\rm m}$, as depicted in the inset of Fig.~1(a). The larger the rate $\gamma_{\rm m}$ and the smaller the time $t_{\rm m}$, the greater the intensity of radiation when the superradiance burst occurs. It is well known that, when the number of emitters is large, it is difficult to numerically solve their dynamic evolution. Therefore, it would be useful if we could characterize the superradiance properties using $g^{(2)}(0)$ as shown in Eq.~(12), which is very easy to obtain as long as the matrix $\boldsymbol \Gamma$ is known.

Ref.~\cite{masson22nc} discussed in detail how the superradiance properties of emitter arrays of different dimensions in free space change with the emitter-emitter distance. Here, by solving the dynamic evolution of the few-emitter system represented with Eq.~(1), we extract the maximum total emission rate $\gamma_{\rm m}$ and the corresponding time $t_{\rm m}$ in the case of all-to-all interactions [i.e., the decoherence matrix is represented in Eq.~(15)], and plot their relationship with the second-order correlation function $g^{(2)}(0)$, as shown in Fig.~1. Different curves in Figs.~1 (a) and 1(b) represent the cases of emitter number $N=3,4,5,6$. Different $g^{(2)}(0)$ in the same curve is realized by changing the parameter $\alpha$ in Eq.~(15) from 0 to 1, which corresponds to that the interaction strength ranges from zero to maximum.

As illustrated in Fig.~1, when $g^{(2)}(0)$ is less than 1, superradiance will not occur. In this case, the emission rate decays exponentially with time, so $t_{\rm m}$ is always 0 and $\gamma_{\rm m}$ is always 1 [all the emission rates in this work are normalized by ${\rm tr}({\boldsymbol \Gamma})$]. As $g^{(2)}(0)$ exceeds 1, $\gamma_{\rm m}$ begins to be greater than 1, and $t_{\rm m}$ begins to be greater than 0, which means superradiance occurs. As $g^{(2)}(0)$ further increases, for a fixed $N$, both $\gamma_{\rm m}$ and $t_{\rm m}$ gradually increase. For the $\gamma_{\rm m}$, the relationship between it and the $g^{(2)}(0)$ is roughly linear when $g^{(2)}(0)$ is large, while the corresponding $t_{\rm m}$ gradually tends to a stable value. 

The results in Fig.~1 have many-fold meanings for the all-to-all case. Firstly, it verifies that $g^{(2)}(0)>1$ is a necessary and sufficient condition to indicate superradiance (this determination of sufficiency and necessity holds true for any complex emitter environment, which can be determined by taking a large number of random decoherence matrices and is not shown here). Secondly, it points out how to efficiently increase the intensity of superradiance emission in the case of all-to-all interaction, that is, increase the number of emitters or enhance the incoherent coupling $\alpha$. Moreover, we can observe that when $g^{(2)}(0)$ is large enough, there is almost a linear relationship between maximum emission rate $\gamma_{\rm m}$ and $g^{(2)}(0)$. And as $N$ increases, the slope of the straight line also increases. Therefore, the results in Fig.~1 can be used to approximately predict the maximum emission rate when $N$ is very large.

\section{Attaining large superradiance using ENZ structures}
As indicated in Refs.~\cite{masson20prl,masson22nc,lopez23prl}, superradiance cannot occur with only nearest-neighbor coupling, and the longer the interaction range, the larger the correlation $g^{(2)}(0)$ and the emission intensity. Therefore, to reach large $g^{(2)}(0)$, we need to build a system of long-distance interacting emitters in order to pursue increasing emission intensity. This all-to-all interaction cannot be achieved in free space unless the density of emitters increases to infinity, as we know that the coupling between regularly arranged emitter arrays in free space rapidly decays with increasing distance. 


In recent years, the emergence of NZI materials \cite{Ziolkowski04pre,alu07prb,Engheta13sci,maas13np,lib17np} has made this all-to-all interaction scenario possible. Among NZI materials, ENZ structures have draw a lot of attention because they support infinitely long effective wavelengths without altering the phase of the electric field in space. Owing to this remarkable characteristic, emitters in these systems are increasingly focusing on long-range \cite{li19prb} and location-insensitive interactions \cite{zhu23oe}. In this section, we focus on the superradiance properties of two popular ENZ structures: the plasmonic ENZ waveguide and the dielectric ENZ photonic crystal.

\subsection{Superradiance of emitters in plasmonic ENZ waveguides}
In the spatial domain range of hundreds of nanometers, the ENZ waveguide naturally provides a uniform dissipative coupling between any two emitters. The periodic-arranged slit waveguide cell (with grating period $a=b=400$~nm) is shown in the upper panel of Fig.~2(a). The medium surrounding the slit is the metal Ag, whose permittivity is adapted from experiment \cite{john72prb}. The silt is composed of dielectric material with permittivity $\epsilon=2.2$. The emitters are inserted in the central line of the waveguide, as shown in the panel, where the arrows represent the orientation of the dipole moment of emitters. The slit length, width and height are taken as $l=1$~$\rm{\mu m}$, $w=200$~nm, and $h=40$~nm, respectively. 

In this work, we set all the dipole moments of emitters along the $y$-axis. According to Refs.~\cite{alu08prb,argy14ol}, the effective refractive index can be calculated using the dispersion equation of a rectangular waveguide, because the modal distribution of present waveguide is very similar to that of quasi-TE$_{10}$ mode \cite{alu08prb}. The real and imaginary parts of the effective index $n_{\rm eff}$ are presented in Fig.~2(b) with solid and dashed lines, respectively. The near-zero index can be observed in the frequency range from $290$ to $300$~THz. This near-zero index has great potential in achieving long-range interaction between emitters and thus emission burst as described in Sec.~II. At the same time, the emission intensity is proportional to the decay rate of a single emitter. In the inset of Fig.~2(b), we plot the enhancement of spontaneous decay of a single emitter located in the center of the slit waveguide, $\gamma_{\rm center}/\gamma_0$, with $\gamma_0$ being the spontaneous decay rate of the emitter in vacuum. Obviously, this single-emitter decay rate has a sharp peak around the frequency $f=295$~THz, with a more than 800-fold enhancement compared to free space. Therefore, the large spontaneous emission rate and near-zero index make the ENZ waveguide an excellent candidate for achieving photon emission burst.

\begin{figure}[t]
	\epsfig{figure=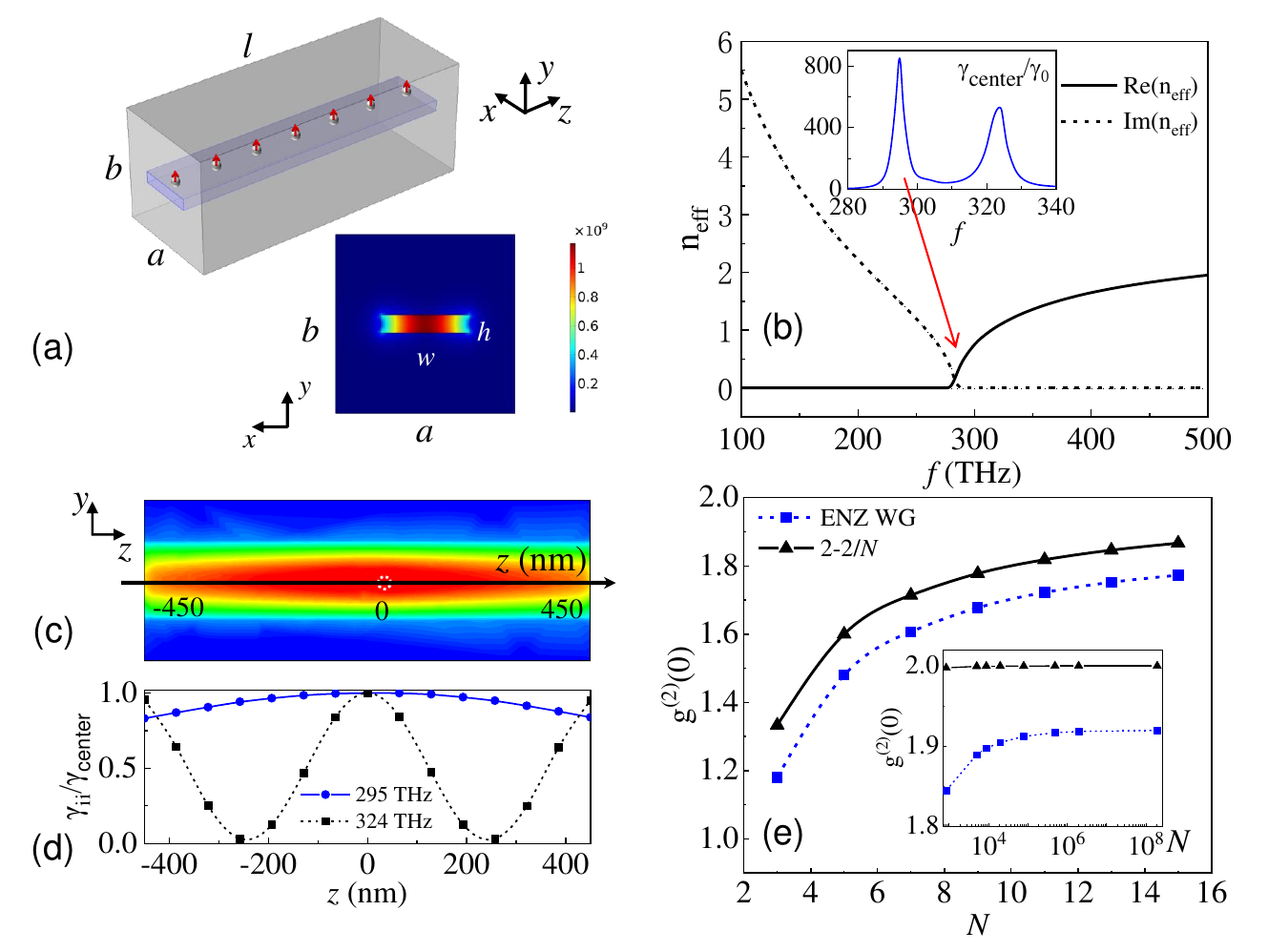,width=0.5\textwidth}
	\caption{(Color online) (a) Upper panel: Schematic diagram of an ENZ waveguide and the emitters inserted in it. Lower panel: The field pattern in the $x-y$ plane of a central emitter with ENZ transition frequency $f=295$~THz. (b) Effective refractive index in the slit as a function of frequency. Inset: The enhancement of single-emitter spontaneous decay rate as a function of frequency. (c) The field distribution in the $y-z$ plane of a single emitter located at the center of the slit with $f=295$~THz. (d) Blue circles and black squares: The normalized spontaneous decay rate of an emitter chain in the central line of the waveguide with frequency $f=295$~THz and $324$~THz, respectively. (e) Blue squares: The dependence of the calculated second-order correlation function $g^{(2)}(0)$ on the emitter number $N$. Black triangles: The corresponding ideal Dicke superradiance as a comparison. }
\end{figure}

Here we first briefly describe how to calculate the dissipative coupling parameters $\gamma_{ij}$ that are necessary for the superradiance indicator $g^{(2)}(0)$. It can be observed that the key to calculate the two parameters is calculating Green's tensor, according to Eq.~(4). For the identical orientation of the dipole moments of emitters, the Green's function $\stackrel{\leftrightarrow}{G}\left(\vec{r}_i, \vec{r}_j, \omega\right)$ can be obtained by calculating the electric field emitted by a point dipole $\vec{\mu}_j$ located at $\vec{r}_j$, i.e.
\begin{equation}
	\stackrel{\leftrightarrow}{G}\left(\vec{r}_i, \vec{r}_j, \omega\right)\cdot\vec{\mu }_j=-\vec{E}\left(\vec{r}_i\right)|_{\vec{\mu}_j},
\end{equation}
where $\vec{E}\left(\vec{r}_i\right)|_{\vec{\mu}_j}$ is the electric field at $\vec{r}=\vec{r}_i$, which is the field emitted from the emitter located at $\vec{r}=\vec{r}_j$ with dipole moment $\vec{\mu}_j$. For the single-emitter spontaneous decay rate $\gamma_{ii}$, we only need to calculate the electric field it feels by emitting from itself. 

In the lower panel of Fig.~2(a) and Fig.~2(c), we present the amplitude of electric field of a unit dipole with transition frequency $f=295$~THz located at the center of the slit in the $x$-$y$ and $y$-$z$ planes, respectively, and all the electric field simulations in this work (except for the vacuum case, which is readily at hand with the analytic formula) are performed using COMSOL multiphysics. The black arrow in Fig.~2(c) represents the line where the emitters are located, and the range of $z$-axis is from -450 to 450~nm. According to the two field patterns we know that, the field emitted from the center emitter is confined in the dielectric slit and distributed almost uniformly, especially near the central line of the slit waveguide. Even though, it is obvious that there is a decrease from the center of the slit to both sides of the waveguide. Therefore, naturally, when the emitters are lined up on the central axis of the slit [as shown in the upper panel of Fig.~2(a)], the spontaneous emission rates of emitters at different positions will be different. We plot the relationship between the normalized spontaneous emission rate of emitters and their positions (represented by $z$) with blue circles in Fig.~2(d). It can be seen from the figure that the spontaneous emission rate of emitters at ENZ frequency gradually and symmetrically decreases from the center to both ends, reducing to about 80$\%$ of the center at the location of -450 or 450 nm. For other resonance frequencies (non-ENZ frequencies), the single-emitter decay rate can oscillate violently with the emitter position. For example, the case for resonance frequency $f=324$~THz [corresponding to the second resonance peak in the inset shown in Fig.~2(b)] is plotted with black squares in Fig.~2(d), where the single-emitter decay rate oscillates like a sinusoidal function as the emitter position varies.

\begin{figure}[t]
	\epsfig{figure=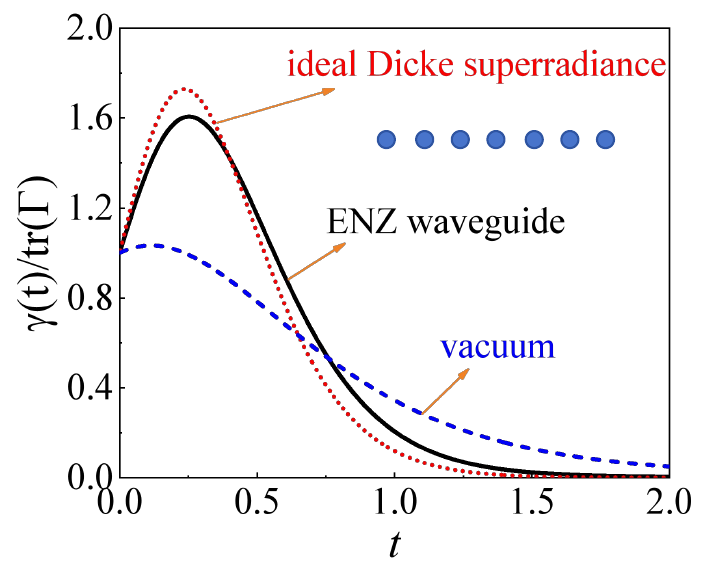,width=0.3\textwidth}
	\caption{(Color online) Total normalized emission rate $\gamma(t)/{\rm tr({\boldsymbol \Gamma})}$ versus time $t$ with three cases: The 7 emitters are in a line in the ENZ waveguide (black solid line), idea Dicke superradiance (red dotted line), and in the vacuum (blue dashed line). The emitters have the same spatial arrangement in three cases, as indicated in the inset. }
\end{figure}

To calculate the correlation function $g^{(2)}(0)$ to judge whether the superradiance occurs or not, we can use Eq.~(12) for all the coupling cases. The blue square-dashed line in Fig.~2(e) shows the dependence between the $g^{(2)}(0)$ of uniformly arranged emitters and the number of emitters $N$ that we calculated. The black triangle line is the case for ideal Dicke superradiance with $g^{(2)}(0)=2-2/N$ as a comparison. The detailed data of decoherence matrix for a 15-emitter system can be found in Supplemental Material. It can be seen that $g^{(2)}(0)$ of emitters in the ENZ waveguide is close and almost parallel to the ideal Dicke superradiance under small emitter numbers. We can get $g^{(2)}(0)$ when the number of emitters is uniformly raised on the central axis of the slit by applying numerical interpolation because the field varies with the emitter position at the ENZ frequency in a fairly smooth manner. When $N$ is large enough, as shown in the inset of Fig.~2(e), $g^{(2)}(0)$ can reach about 1.9, which is very close to the maximum value of 2.

To demonstrate the similarity between the emission dynamics of emitters in ENZ waveguide and the ideal Dicke superradiance, we plot the evolution of the normalized total emission rate over time for the case of seven emitters in Fig.~3. The black solid and the red dotted lines in the figure represent the ENZ waveguide and the ideal Dicke superradiance, respectively. It can be seen that they have very close dynamic curves and similar maximum emission rates, which shows that the superradiance in such an plasmonic ENZ waveguide is very close to Dicke superradiance. As a comparison, we also plot the case that emitters in vacuum and have the same spatial arrangement. In vacuum, the dissipative coupling between two emitters (both of their dipole moments are perpendicular to the line that contains two emitters) can be calculated as
\begin{equation}
	\frac{{\gamma}_{ij}}{\gamma_{0}}
	= \frac{3}{2 k_0 R}\left[ {\rm sin}(k_0R)+\frac{{\rm cos}(k_0R)}{k_0R}-\frac{{\rm sin}(k_0R)}{k_0^2R^2} \right],  
\end{equation}
where $k_0$ is the wavevector in vacuum, and $R$ is the distance between two emitters. A brief derivation of the formula can be found in Supplemental Material. The calculated result is presented with a blue dashed line in Fig.~3, where we can find that, although the superradiance also occurs at such a small distance, the maximum emission rate is much smaller than in the ENZ and ideal Dicke cases.



\subsection{Superradiance of emitters in dielectric ENZ photonic crystal}

According to Sec.~III~A we know that long-range interactions between emitters and, thus, near-ideal superradiance can be attained via plasmonic waveguide with near-zero index, and the spatial location requirements for emitters need not to be very tight. Nevertheless, there are two drawbacks to this plasmonic waveguide system for increasing superradiance. On the one hand, single-emitter decay and long-range interactions are inevitably impacted by dissipation in environment because of the significant loss in plasmonic systems; consequently, the range of emitter-emitter interactions is lowered. On the other hand, there are additional stringent restrictions for the emitter frequency in plasmonic systems. For instance, as illustrated in Fig.~2(b), the ENZ and resonance frequencies need to be close to 295~THz. This also narrows the system's application scope. These drawbacks can be overcame by all-dielectric ENZ photonic crystals \cite{moitra13natp,liberal17pnas,vulis17oe,mello22apl} since dielectric materials naturally prevent dissipation and scaling features are inherent to photonic crystals. Therefore, the ENZ photonic crystal has the capability to attain interactions over longer distances and consequently, superradiance throughout a broader spatial domain. In addition, the photonic crystal's scaling features allow for flexible modulation of the superradiance frequency.

\begin{figure}[t]
	\epsfig{figure=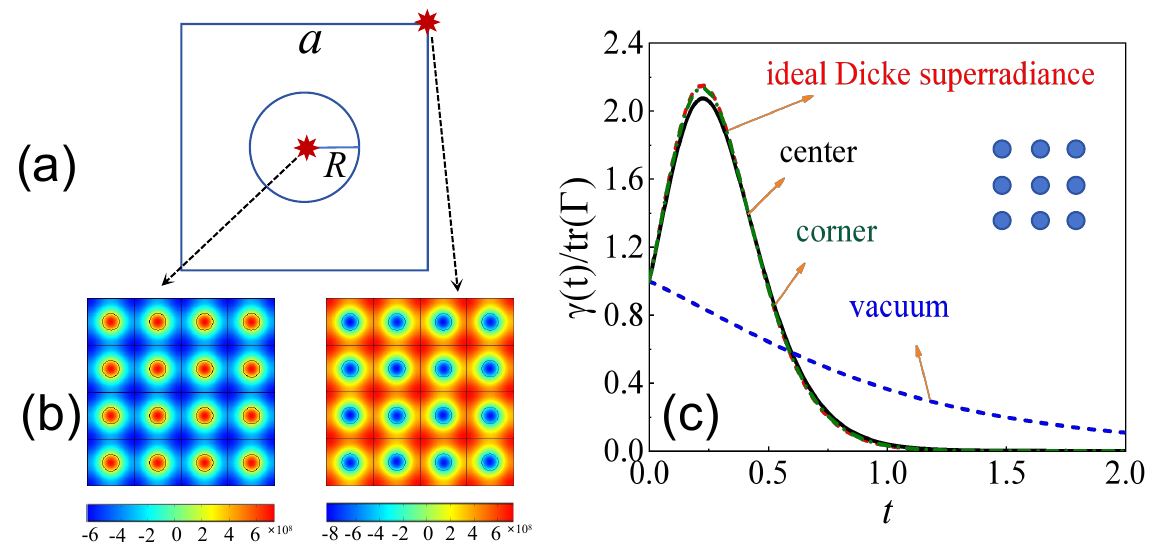,width=0.5\textwidth}
	\caption{(Color online) (a) The schematic diagram of photonic crystal cell. (b) Field patterns of single emitter located at the center of the cylinder and the corner of the cell, respectively. (c) Total normalized emission rate $\gamma(t)/{\rm tr({\boldmath \Gamma})}$ versus time $t$ with four cases: The nine emitters are arranged with a 3$\times$3 array at the center of cylinders (black solid line), at the corners of cells (green dot-dashed line), ideal Dicke superradiance (red dotted line) and in the vacuum (blue dashed line). The emitters have the same spatial arrangement in four cases, as depicted in the inset. }
\end{figure}

Here we choose a diamond cylinder square photonic crystal whose ENZ frequency is within the visible light range \cite{mello22apl}, to demonstrate the efficiency for near-ideal and large-domain superradiance. The unit cell of this photonic crystal is depicted in Fig.~4(a), where the lattice constant $a=480.3$~nm, the radius of the cylinder $R=85.3$~nm and the refractive index of the cylinder is $n=2.41$, and the surrounding is air. The energy band diagram of the photonic crystal and the effective index of the structure can be found in Supplemental Material. Two panels of Fig.~4(b) show respectively the distribution of the electric field generated in space when an emitter is located at the center of the cylinder or the corner of the cell [see the two red stars in Fig.~4(a)]. Here we calculate the field distribution of a 71$\times$71 cylindrical array, but in order to see the details clearly, only the 4$\times$4 array is shown here. According to Fig.~4(b), regardless of whether the source is at the center of the cylinder or the corner of the cell, the field distribution in the space changes periodically with the lattice. This periodic field distribution originates from the near-zero effective index. In addition, we observe that the dissipative coupling along the straight line at the corner is more uniform along the straight line at the center of the cylinder, because the electric field inside and outside the cylinder will change drastically. In other words, if we randomly insert emitters in this ENZ photonic crystal, in order to obtain a more uniform decoherence matrix, it is better to place the emitters in the space around the cylinders rather than in them. Detailed information can be found in the Supplemental Material.

We calculate the decoherence matrix of nine emitters located in the center and corner of the cylinders arranged as a 3$\times$3 array, as shown in the inset of Fig.~4(a). The calculated data can be found in Supplemental Material. The second-order correlation can be easily calculated as $g^{(2)}(0)=1.725$ and $1.768$, respectively, which are very close to that for an ideal case with $g^{(2)}(0)=1.7778$. The dynamics of total emission rate in these two cases are plotted with solid black and green dot-dashed lines in Fig.~4(c), where the red dotted and blue dashed lines are the ideal Dicke superradiance and the case in vacuum, respectively. According to Fig.~4(c), the emission dynamics of nine emitters embedded in ENZ photonic crystal is very close to the ideal Dicke superradiance, and the superradiance even does not occur in the vacuum under the same spatial arrangement. Therefore, the ENZ photonic crystal can provide near-ideal all-to-all interaction in a more extended spacial domain.



\section{Role of quantum coherence in photon emission dynamics of all-inverted emitters}

\begin{figure}[t]
	\epsfig{figure=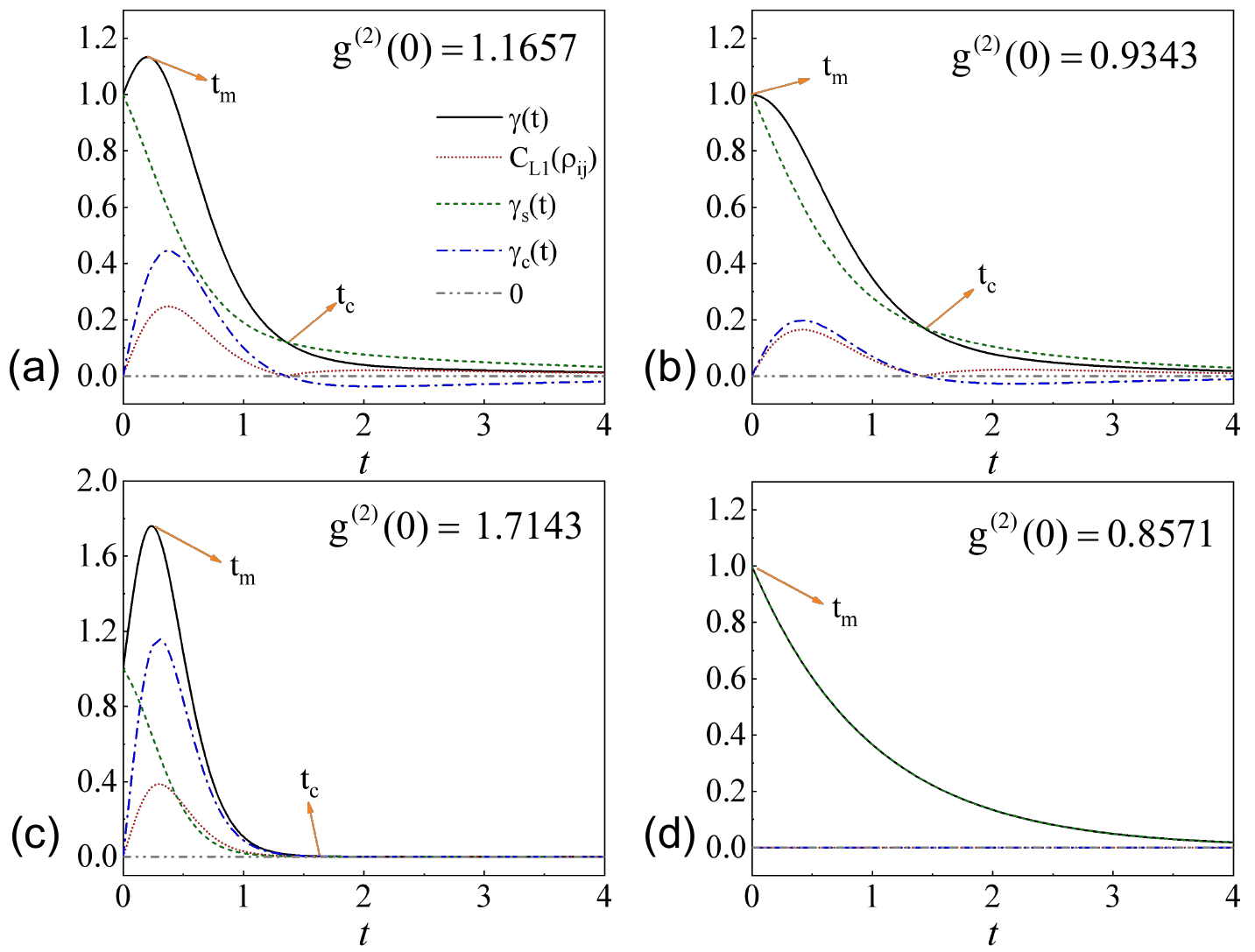,width=0.5\textwidth}
	\caption{(Color online) Role of quantum coherence in emission dynamics. The black solid lines are the total emission rate $\gamma(t)$, green dashed and red dotted lines represent the single-emitter and collective emission rates $\gamma_{\rm s}(t)$ and $\gamma_{\rm c}(t)$, respectively. The blue dash-dotted line is the two-emitter $l_1$-norm coherence $C_{L1}$. The gray dot-dot-dashed line is the reference zero line. The $\gamma(t)$, $\gamma_{\rm s}(t)$ and $\gamma_{\rm c}(t)$ are all normalized with the emitter number $N=7$. The diagonal elements of $\boldsymbol \Gamma$ matrix are 1 and the non-diagonal elements $\alpha$ are taken as (a) 0.6, (b) 0.3, (c) 1.0 and (d) 0.0. }
\end{figure}

In Sec.~II we derived the superradiance condition using the decoherence matrix, which concerns the dynamics of emission at the early stage of the whole emission process. We now turn to consider the later emission dynamics and explore the role of correlations in the dynamics of collective emission. Note that here we use the term collective emission instead of superradiance. This is because, as we will see in this Section, the appearance of superradiance must be accompanied by collective emission, but superradiance does not necessarily occur when collective emission occurs.
 
It can be seen from Eq.~(9) that the dissipative coupling $\gamma_{ij}$ is always beneficial to the increase of the total radiation rate at the early stage of the emission process, but according to Eq.~(7), the collective contribution $\gamma_{\rm c}(t)$ may be negative (because the average of $\hat{\sigma}_{i}^{\dagger}\hat{\sigma}_{j}$ at time $t$ can be negative), which means that in the later stage of emission, correlation may be harmful to the total spontaneous radiation rate. In this case, there must be a crossover between total emission $\gamma(t)$ and single-emitter emission $\gamma_{\rm s}(t)$, as shown in the following analysis in the detailed examples.

We still first consider the all-to-all case, that is, the decoherence matrix is homogeneous as shown in Eq.~(15). We plot the total emission rates as a function of time in Fig.~2. The number of emitters is fixed at $N=7$, and $\alpha=$0.6, 0.3, 1.0 and 0.0 correspond to four panels, respectively. According to Eq.~(16), it is easy to calculate that $g^{(2)}(0)=$1.1657, 0.9343, 1.7143 and 0.8571 in four cases. In four panels, the black solid lines represent the total emission rate $\gamma(t)$, the green dashed and blue dot-dashed lines represent single-emitter and collective emission parts $\gamma_{\rm s}(t)$ and $\gamma_{\rm c}(t)$, respectively. As marked in Fig.~5, $t_{\rm m}$ represents the time that corresponds to the maximum emission rate, and $t_{\rm c}$ is the crossover between the total decay rate $\gamma(t)$ and single-emitter decay rate $\gamma_{\rm s}(t)$. If $t_{\rm m}=0$, as shown in Figs.~5(b) and 5(d), we have $g^{(2)}(0)<1$, and there is no superradiance. Otherwise, as shown in Figs.~5(a) and 5(c), the second-order correlation $g^{(2)}(0)>1$ and the nonzero $t_{\rm m}$ indicate superradiance. In Fig.~5(b), during the time period from 0 to $t_{\rm c}$, although there is no emission burst, the total spontaneous emission is still greater than the single-emitter spontaneous emission, that is, collective emission plays a positive role in the emission process. This can be demonstrated through the dynamics of the collective part of emission $\gamma_{\rm c}(t)$. Figs.~5(a) and (b) present an obvious cross between the $\gamma(t)$ and $\gamma_{\rm s}(t)$, where we can find that the crossover corresponds exactly to the time that $\gamma_{\rm c}(t)$ turns from positive to negative. Therefore, no matter whether superradiance occurs or not, $\gamma_{\rm c}(t)$ is always positive when $t<t_{\rm c}$, and after $t_{\rm c}$, the collective effect has a negative effect on the total photon emission. 

Although $\gamma_{\rm c}$ is the collective effect caused by correlations between emitters, it is not a genuine quantum resource because it can be negative. Then it is interesting to consider the relationship between quantum resources and collective emission. Under the scenario of all-to-all interactions, all emitters have the same environment, therefore, here we use the $l_1$-norm quantum coherence \cite{ple14prl} of the reduced density matrix of any two emitters $C_{L1}$ to measure the quantum resource between emitters, which is equal to the sum of the modulus of the off-diagonal elements
\begin{equation}
	C_{L1}(\rho)=\sum_{i\neq j}|\rho_{ij}|,
\end{equation}
where $\rho_{ij}$s are the elements of the two-emitter reduced density matrix $\rho$. 

The quantum coherence in four cases is plotted with red dashed lines in Fig.~5. As shown in the four panels, within the time period from 0 to $t_{\rm c}$, quantum coherence is always synchronous with the collective part $\gamma_{\rm c}$, that is, they increase from 0, then reach the maximum, and at last decrease to 0 at the same time. But after $t_{\rm c}$, the quantum coherence suddenly recovers from zero, which is different from $\gamma_{\rm c}$ that smoothly decreases to a negative value. This illustrates that non-zero quantum correlations can also be harmful to collective emission. In addition, the synchrony of quantum coherence and collective radiation provides us with a way to determine when collective radiation ends. As shown in Eq.~(7), collective emission requires measurement of all emitter-emitter correlations. However, by measuring the quantum coherence between one pair of emitters, we can determine when the collective emission ends.

\begin{figure}[t]
	\epsfig{figure=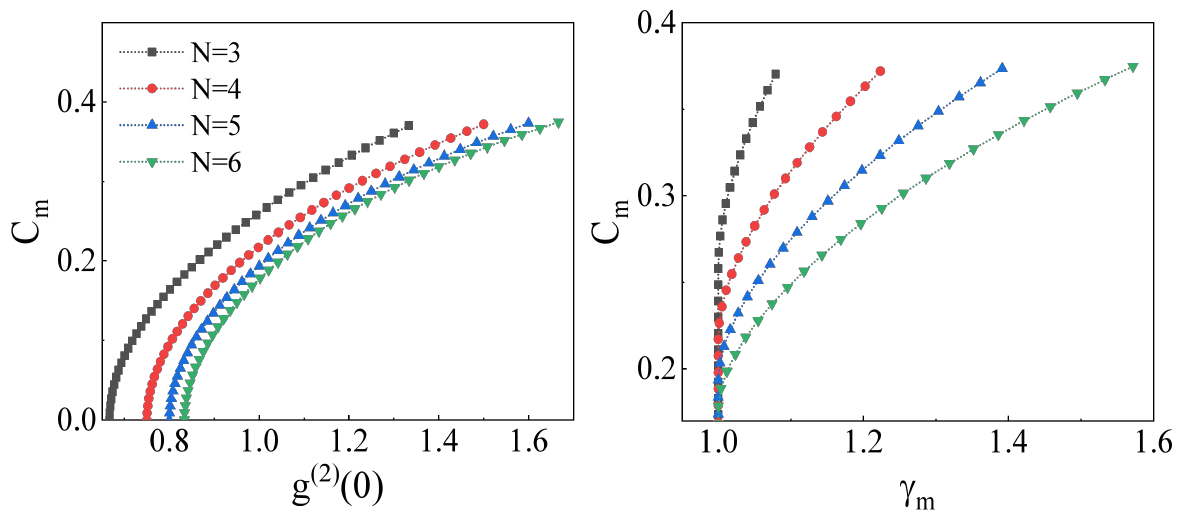,width=0.5\textwidth}
	\caption{(Color online) Dependence of maximum two-emitter quantum coherence $C_{\rm m}$ on the (a) second-order correlation $g^{(2)}(0)$ and (b) maximum emission rate $\gamma_{\rm m}$, with different emitter numbers $N$. }
\end{figure}

More importantly, according to Fig.~5, we find that the maximum value of quantum coherence $C_{\rm m}$ is strongly dependent on the second-order correlation function $g^{(2)}(0)$ and the maximum emission rate $\gamma_{\rm m}$. By solving the dynamics of few-emitter systems, we extract the maximum information of quantum coherence and plot its dependence on $g^{(2)}(0)$ and $\gamma_{\rm m}$ in Fig.~6(a) and 6(b), respectively. According to Fig.~6(a) we find that the maximum $C_{\rm m}$ increases exponentially with the increase of $g^{(2)}(0)$ when $g^{(2)}(0)$ is small, and then increases almost linearly with $g^{(2)}(0)$. Therefore, the maximum quantum coherence can be inferred from $g^{(2)}(0)$, regardless of whether superradiance occurs or not.

What is more interesting is that, according to the dependence of $C_{\rm m}$ on the $\gamma_{\rm m}$ as shown in Fig.~6(b), when superradiance occurs, the quantum correlation in the system can be inferred just by measuring the maximum value of the total emission rate of the multi-emitter system without any quantum operator measurement. Since the measurement of total radiation intensity is direct and simple, this result provides a new suggestion for measuring the maximum value of quantum correlation in multi-emitter systems.

\section{Summary}
In conclusion, we have developed a method to predict the occurrence of superradiance in multi-emitter systems within complex electromagnetic environments by utilizing the derived second-order correlation at the initial time. This approach involves analyzing the $N$-dimensional decoherence matrix exclusively. It is demonstrated that, the emission of emitters in both structures can be very close to the ideal Dicke superradiance, enhancing the potential for achieving near-ideal superradiance in expanded domains using near-zero-index materials. But both the plasmonic and photonic crystal systems possess distinct advantages and disadvantages. Plasmonic systems, despite exhibiting unavoidable losses, excel at achieving location-insensitive long-distance interactions. On the other hand, the photonic crystal system, while avoiding losses, enables the attainment of more ideal superradiance within a larger spatial domain. However, it imposes stricter constraints on the positioning of emitters. Furthermore, through an examination of single-emitter and collective emission in all-to-all interaction systems, we observed a strong synchronization between pairwise quantum coherence and collective emission. Specifically, quantum coherence consistently contributes to the enhancement of radiation rates during collective emission. However, once collective emission ceases, the subsequent recovery of non-zero coherence becomes detrimental to the emission rate. Additionally, our analysis reveals a high correlation between the maximum quantum coherence and the maximum emission rate when superradiance occurs. This suggests that the quantum coherence within the system may be estimated by measuring the readily accessible emission rate. These findings contribute to the advancement of both theoretical understanding in the field of superradiance and the application of photon emission dynamics in quantum information processing.

\begin{acknowledgments}
	This work was supported by the CRF of Hong Kong (C6009-20G), the NSF-China (Grant Nos. 11575051, 11904078), and Hebei NSF (Grant Nos. A2019205266, A2021205020). JR was also funded by the project of China Postdoctoral Science Foundation (Grant No. 2020M670683).
\end{acknowledgments}



\setcounter{equation}{0}
\setcounter{figure}{0}
\onecolumngrid
\newpage
\section*{Supplemental Material}

\subsection*{I. The derivation of the second-order correlation by using the quantum jump operator method}
In a weakly coupled environment, after using the Born-Markov and rotating-wave approximations, and tracing out the environment, the Lindblad master equation of the emitter system can be written as \cite{dung02pra,gon11prl,ren19njp}
\begin{equation}
	\frac{\partial \rho}{\partial t}=\frac{i}{\hbar}[\rho, H]+\frac{1}{2} \sum_{i, j} \gamma_{i j}\left(2 \hat{\sigma}_i \rho \hat{\sigma}_j^{\dagger}-\rho \hat{\sigma}_i^{\dagger} \hat{\sigma}_j-\hat{\sigma}_i^{\dagger} \hat{\sigma}_j \rho\right),
\end{equation}
where $\hat{\sigma}_i^{\dagger}$ and $\hat{\sigma}_i$ are the raising and lowering operators of the $i$th emitter, respectively. When dealing with collective spontaneous decay, it is convenient to recast the spin operators of single emitters into the collective jump operator $\left\lbrace \hat{\mathcal{O}}_{\nu}\right\rbrace $, and the Lindblad equation can be written as
\begin{equation}
	\frac{\partial \rho}{\partial t}=\frac{i}{\hbar}[\rho, H]+\frac{1}{2} \sum_{\nu} \Gamma_{\nu}\left(2 \hat{\mathcal{O}}_{\nu} \rho \hat{\mathcal{O}}_{\nu}^{\dagger}-\rho \hat{\mathcal{O}}_{\nu}^{\dagger} \hat{\mathcal{O}}_{\nu}-\hat{\mathcal{O}}_{\nu}^{\dagger} \hat{\mathcal{O}}_{\nu} \rho\right),
\end{equation}
with $\left\lbrace \Gamma_{\nu}\right\rbrace $ being the eigenvalues of decoherence matrix $\boldsymbol \Gamma$ which has the form
\begin{equation}
	\begin{split}
		&\boldsymbol \Gamma=
		\begin{pmatrix} \gamma_{11} & \gamma_{12} & \cdots & \gamma_{1N}  \\ \gamma_{21} & \gamma_{22} & \cdots & \gamma_{2N} \\ \vdots & \vdots & \ddots & \vdots \\ \gamma_{N1} & \gamma_{N2} & \cdots & \gamma_{NN} \end{pmatrix},
	\end{split}
\end{equation}
where the dissipative coupling $\gamma_{ij}$ is the dissipative coupling between the $i$th and $i$th emitters, which can be calculated using electromagnetic Green's tensor, as shown in Eq.~(4) of the main text. The collective $\nu$-jump operator $\hat{\mathcal{O}}_{\nu}$ can be written as
\begin{equation}
	\hat{\mathcal{O}}_{\nu}=\sum_{i=1}^{N}\alpha_{\nu,i}\hat{\sigma}_{i}, \hat{\mathcal{O}}_{\nu}^{\dagger}=\sum_{i=1}^{N}\alpha_{\nu,i}^{*}\hat{\sigma}_{i}^{\dagger}
\end{equation}
and $\left( \alpha_{\nu,1},\alpha_{\nu,2},\cdots,\alpha_{\nu,N}\right) ^T$ is the normalized eigenvector of matrix $\boldsymbol \Gamma$ corresponds to the eigenvalue $\Gamma_{\nu}$, thus we have
\begin{equation}
	\sum_{i=1}^{N}\alpha_{\nu,i}^{*}\alpha_{\mu,i}=\delta_{\nu\mu}.
\end{equation}

In this case, the matrix elements of $\boldsymbol \Gamma$ showed in Eq.~(3) is
\begin{equation}
	\gamma_{ij}=\sum\limits_{\nu=1}^N \Gamma_{\nu}\alpha_{\nu,i}\alpha_{\nu,j}^*,
\end{equation}
and naturally, $\gamma_{ii}=\sum\limits_{\nu=1}^N \Gamma_{\nu}|\alpha_{\nu,i}|^2$.

The superradiance can be captured by a second-order correlation function
\begin{equation}
	g^{(2)}(0)=	\frac{\sum\limits_{\nu,\mu=1}^N\boldsymbol \Gamma_\nu \boldsymbol \Gamma_\mu\\
		\left\langle \hat{\mathcal{O}}_{\nu}^{\dagger} \hat{\mathcal{O}}_{\mu}^{\dagger} \hat{\mathcal{O}}_{\mu} \hat{\mathcal{O}}_{\nu} \right\rangle }{\left( \sum\limits_{\nu=1}^N\boldsymbol \Gamma_\nu \left\langle \hat{\mathcal{O}}_{\nu}^{\dagger} \hat{\mathcal{O}}_{\nu} \right\rangle\right) ^2},
\end{equation}
where the average is taken on the fully inverted state $|e\rangle^{\otimes N}$ at the start time. Substituting the $\nu$-jump operator showed in Eq.~(4) into the formula above, we have
\begin{equation}
	g^{(2)}(0)=\frac{\sum\limits_{\nu,\mu=1}^{N}\Gamma_{\nu}\Gamma_{\mu}\sum\limits_{i,j,l,m=1}^{N}\alpha_{\nu,i}^{*}\alpha_{\mu,j}^{*}\alpha_{\mu,l}\alpha_{\nu,m}\Big\langle\hat{\sigma}_{i}^{\dagger}\hat{\sigma}_{j}^{\dagger}\hat{\sigma}_{l}\hat{\sigma}_{m}\Big\rangle}{\left(\sum\limits_{\nu=1}^{N}\Gamma_{\nu}\sum\limits_{i,j=1}^{N}\alpha_{\nu,i}^{*}\alpha_{\nu,j}\Big\langle\hat{\sigma}_{i}^{\dagger}\hat{\sigma}_{j}\Big\rangle\right)^{2}}.
\end{equation}
For the fully inverted initial state, we have $\left\langle \hat{\sigma}_{i}^{\dagger}\hat{\sigma}_{j}\right\rangle =\delta_{i j}$ and $\left\langle\hat{\sigma}_{i}^{\dagger}\hat{\sigma}_{j}^{\dagger}\hat{\sigma}_{l}\hat{\sigma}_{m}\right\rangle=\left(\delta_{i m}\delta_{j l}+\delta_{i l}\delta_{j m}\right)\left(1-\delta_{i j}\right)$. Therefore, the second-order correlation function is
\begin{align}
	g^{(2)}(0) &= \frac{\sum\limits_{\nu,\mu=1}^{N}\Gamma_{\nu}\Gamma_{\mu}\biggl(\sum\limits_{i,j=1}^{N}|\alpha_{\nu,i}|^{2}|\alpha_{\mu,j}|^{2}\ +\ \sum\limits_{i,j=1}^{N}\alpha_{\nu,i}^{*}\alpha_{\mu,j}^{*}\alpha_{\mu,i}\alpha_{\nu,j}\ -\ 2\sum\limits_{i=1}^{N}|\alpha_{\nu,i}|^{2}|\alpha_{\mu,i}|^{2}\biggr)}{\left(\sum\limits_{\nu=1}^{N}\Gamma_{\nu}\sum\limits_{i=1}^{N}|\alpha_{\nu,i}|^{2}\right)^{2}} \nonumber\\
	&= \frac{\sum\limits_{\nu,\mu=1}^{N}\Gamma_{\nu}\Gamma_{\mu}\biggl[( \sum\limits_{i=1}^{N}|\alpha_{\nu,i}|^{2})(\sum\limits_{j=1}^{N}|\alpha_{\mu,j}|^{2})\ +\ (\sum\limits_{i=1}^{N}\alpha_{\nu,i}^{*}\alpha_{\mu,i})(\sum\limits_{j=1}^{N}\alpha_{\mu,j}^{*}\alpha_{\nu,j})\ -\ 2\sum\limits_{i=1}^{N}|\alpha_{\nu,i}|^{2}|\alpha_{\mu,i}|^{2}\biggr]}{\left(\sum\limits_{\nu=1}^{N}\Gamma_{\nu}\sum\limits_{i=1}^{N}|\alpha_{\nu,i}|^{2}\right)^{2}} \nonumber\\
	&= \frac{\sum\limits_{\nu,\mu=1}^{N}\Gamma_{\nu}\Gamma_{\mu}\biggl[1\,+\delta_{\nu\mu}\,-\,\sum\limits_{i=1}^{N}\,2|\alpha_{\nu,i}|^{2}\,|\alpha_{\mu,i}|^{2}\biggr]}{\left(\sum\limits_{\nu=1}^{N}\Gamma_{\nu}\right)^{2}}  \nonumber\\
	&= \frac{\sum\limits_{\nu,\mu=1}^{N}\Gamma_{\nu}\Gamma_{\mu}+\sum\limits_{\nu=1}^{N}\Gamma_{\nu}^2-2\sum_{i=1}^{N}\left(\sum_{\nu=1}^{N}\Gamma_{\nu}|\alpha_{\nu,i}|^{2}\right)\left(\sum_{\mu=1}^{N}\Gamma_{\mu}|\alpha_{\mu,i}|^{2}\right)}{\left(\sum\limits_{\nu=1}^{N}\Gamma_{\nu}\right)^{2}} \nonumber\\
	&= \frac{\sum\limits_{\nu,\mu=1}^{N}\Gamma_{\nu}\Gamma_{\mu}+\sum\limits_{\nu=1}^{N}\Gamma_{\nu}^2-2\sum_{i=1}^{N}\gamma_{ii}^2}{\left(\sum\limits_{\nu=1}^{N}\Gamma_{\nu}\right)^{2}} \nonumber\\
	&= \frac{\left({\rm Tr} \boldsymbol \Gamma\right) ^2+{\rm Tr}\left(\boldsymbol \Gamma^2 \right)-2{\rm Tr}\left(\boldsymbol \Gamma_{\rm d}^2 \right) }{\left({\rm Tr} \boldsymbol \Gamma\right) ^2}.		
\end{align}
with ${\boldsymbol \Gamma}_{\rm d}={\rm diag(\gamma_{11},\gamma_{22},\cdots,\gamma_{NN})}$ being the diagonal matrix of decoherence matrix $\boldsymbol \Gamma$.

\subsection*{II. Non-uniform field induced non-identical decays and couplings}

%

If all the emitters are in the same electromagnetic environment, that is their spontaneous decay are identical, like $\gamma_0$, the $g^{(2)}(0)$ can take the maximum value
\begin{align}
	g^{(2)}(0)&=	1+\frac{\rm{Tr}(\boldsymbol \Gamma^2)}{(\rm{Tr}(\boldsymbol \Gamma))^2}-\frac{2}{N},\nonumber\\
	&=1+{\frac{1}{N}}\left[\mathrm{Var}\left({\frac{\{\Gamma_{\nu}\}}{\gamma_{0}}}\right)-1\right], 
\end{align}
which is the same to the existing results \cite{masson22nc,lopez23prl}, where the free-space condition is assumed. According to Eq.~(10), the superradiance condition $g^{(2)}(0)>1$ corresponds to the variance of $\{\Gamma_{\nu}\}/\gamma_0$ is larger than 1. From the perspective of emitting photons, large variance means that a large number of $\nu$-jump channels are closed, that is, most collective emission modes are dark modes, while the emitter system radiates photons in burst through a small number of collective channels (bright channels).

As indicated in Eq.~(10), when the single emitter emission rate is identity, the $g^{(2)}(0)$ is proportional to the variance of the eigenvalues of matrix $\boldsymbol \Gamma$ for a fixed emitter number $N$. However, when the decay rates of emitters are no longer equal, the $g^{(2)}(0)$ must be calculated using Eq.~(9). Here we want to visually show that Eq.~(10) no longer holds in this case.

In Fig. 1, we plot the relationship between $g^{(2)}(0)$ showed in Eq.~(10) and the variance of eigenvalues of decoherence matrix ${\rm Var}\{\Gamma_\nu\}$ by taking 10000 random decoherence matrix $\boldsymbol \Gamma$ for the two cases. The dimension of the decoherence matrix $\boldsymbol \Gamma$ (namely the number of the emitters) is fixed at $N=1000$. In Fig.~1(a) all the diagonal elements of $\boldsymbol \Gamma$ is taken as unity, and the other elements are random numbers range from 0 to 1. While for Fig.~1(b) all the matrix elements are taken randomly from 0 to 1, other than that the condition $\gamma_{ij}\leq{\rm min}\{\gamma_{ii},\gamma_{jj}\}$. According to Fig.~1(a), $g^{(2)}(0)$ is indeed proportional to ${\rm Var}\{\Gamma_\nu\}$ when single-emitter emission rates are equal. But this relationship is broken when single-emitter emission rates are not equal, as shown in Fig.~1(b).

\begin{figure}
	\epsfig{figure=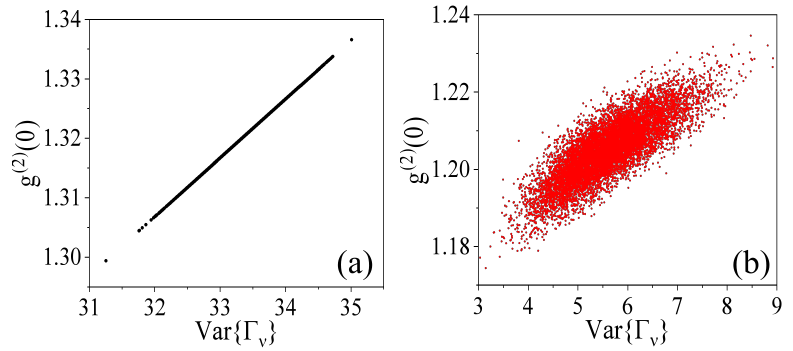,width=0.5\textwidth}
	\caption{(Color online) The second-order correlation $g^{(2)}(0)$ versus variance of eigenvalues of decoherence matrix ${\rm Var}\{\Gamma_\nu\}$ for (a) identical and (b) different single-emitter emission rates. The number of emitter is $N$=1000 and the results of 10000 random decoherence matrices are presented in both panels.}
\end{figure}

Now we verify the sufficiency and necessity of the minimal superradiance condition, by calculating the emission dynamics of random-coupled few-emitter system.

\subsection*{III. The Green's tensor profile of incoherent coupling of two emitters in homogeneous medium}
The compact form of dyadic Green's function in the homogeneous medium is
\begin{equation}
	\overset{\leftrightarrow }{\mathop{G}}\,\left( {{{\vec{r}}}_{i}},{{{\vec{r}}}_{j}},\omega  \right)=\left[\overset{\leftrightarrow }{I}+\frac{1}{k^2}\nabla\nabla \right]\frac{e^{ik|{{\vec{r}}}_{i}-{{\vec{r}}}_{j}|}}{4\pi|{{\vec{r}}}_{i}-{{\vec{r}}}_{j}|},
\end{equation}
where 
\begin{equation}
	\nabla= \frac{\partial}{\partial x}\hat{x}+\frac{\partial}{\partial y}\hat{y}+\frac{\partial}{\partial z}\hat{z},
\end{equation}
and after some calculations its explicit form can be expresses as
\begin{equation}
	\overset{\leftrightarrow }{\mathop{G}}\,\left( {{{\vec{r}}}_{i}},{{{\vec{r}}}_{j}},\omega  \right)=\left\lbrace \left( \frac{3}{k^2R^2}-\frac{3i}{kR}-1 \right)\hat{R}\hat{R}+\left( 1+\frac{i}{kR}-\frac{1}{k^2R^2}\right)\overset{\leftrightarrow }{I}   \right\rbrace \frac{e^{ik R}}{4\pi R},
\end{equation}
with $R=|\vec{r}_i-\vec{r}_j|$, and $\hat{R}=(\vec{r}_i-\vec{r}_j)/|\vec{r}_i-\vec{r}_j|$. 

If the electric dipole moments of two emitters are parallel and perpendicular to the line connecting their positions (that is, perpendicular to $\hat{R}$), then the dissipative interaction between then can be calculated as
\begin{equation}
	\begin{split}
		{{\gamma}_{ij}}&=\frac{2\omega _{0}^{2}}{{{\varepsilon }_{0}}\hbar {{c}^{2}}}\operatorname{Im}\left[ \vec{\mu }_i^{*}\cdot \overset{\leftrightarrow }{\mathop{G}}\,\left( {{{\vec{r}}}_i},{{{\vec{r}}}_j},\omega  \right)\cdot {{{\vec{\mu }}}_j} \right]\\
		&=\frac{2\omega _{0}^{2}}{{\varepsilon _{0}}\hbar {{c}^{2}}} {\rm Im}\left[ 1+\frac{i}{kR}-\frac{1}{k^2R^2}\right] \frac{e^{ik R}}{4\pi R}\\
		&=\frac{2\omega _{0}^{2}}{{{\varepsilon }_{0}}\hbar {{c}^{2}}}\mu^2 \frac{1}{4\pi R}\left[ {\rm sin}(kR)+\frac{{\rm cos}(kR)}{kR}-\frac{{\rm sin}(kR)}{k^2R^2} \right].  
	\end{split}
\end{equation}
The decay rate of an emitter in the homogeneous medium is
\begin{equation}
	\gamma_{ii}=n\left(\frac{\omega^3\mu^2}{3\pi \varepsilon_0 \hbar c^3} \right)=n\gamma_0, 	
\end{equation}
where $\gamma_0=k_0^3\mu^2/(3\pi \varepsilon_0 \hbar)$ is the spontaneous of single emitter in vacuum, $k_0=\omega/c$ is the wavevector in vacuum, and $n$ is the refractive index of the homogeneous medium. Therefore, in a homogeneous medium we have the normalized dissipative coupling
\begin{equation}
	\frac{{\gamma}_{ij}}{\gamma_{ii}}
	= \frac{3}{2nk_0 R}\left[ {\rm sin}(kR)+\frac{{\rm cos}(kR)}{kR}-\frac{{\rm sin}(kR)}{k^2R^2} \right],  
\end{equation}
and in the vacuum,
\begin{equation}
	\frac{{\gamma}_{ij}}{\gamma_{0}}
	= \frac{3}{2 k_0 R}\left[ {\rm sin}(k_0R)+\frac{{\rm cos}(k_0R)}{k_0R}-\frac{{\rm sin}(k_0R)}{k_0^2R^2} \right].  
\end{equation}

%
%

\subsection*{IV. Data of decoherence matrix of emitter in ENZ waveguide and photonic crystal}

\begin{figure}
	\epsfig{figure=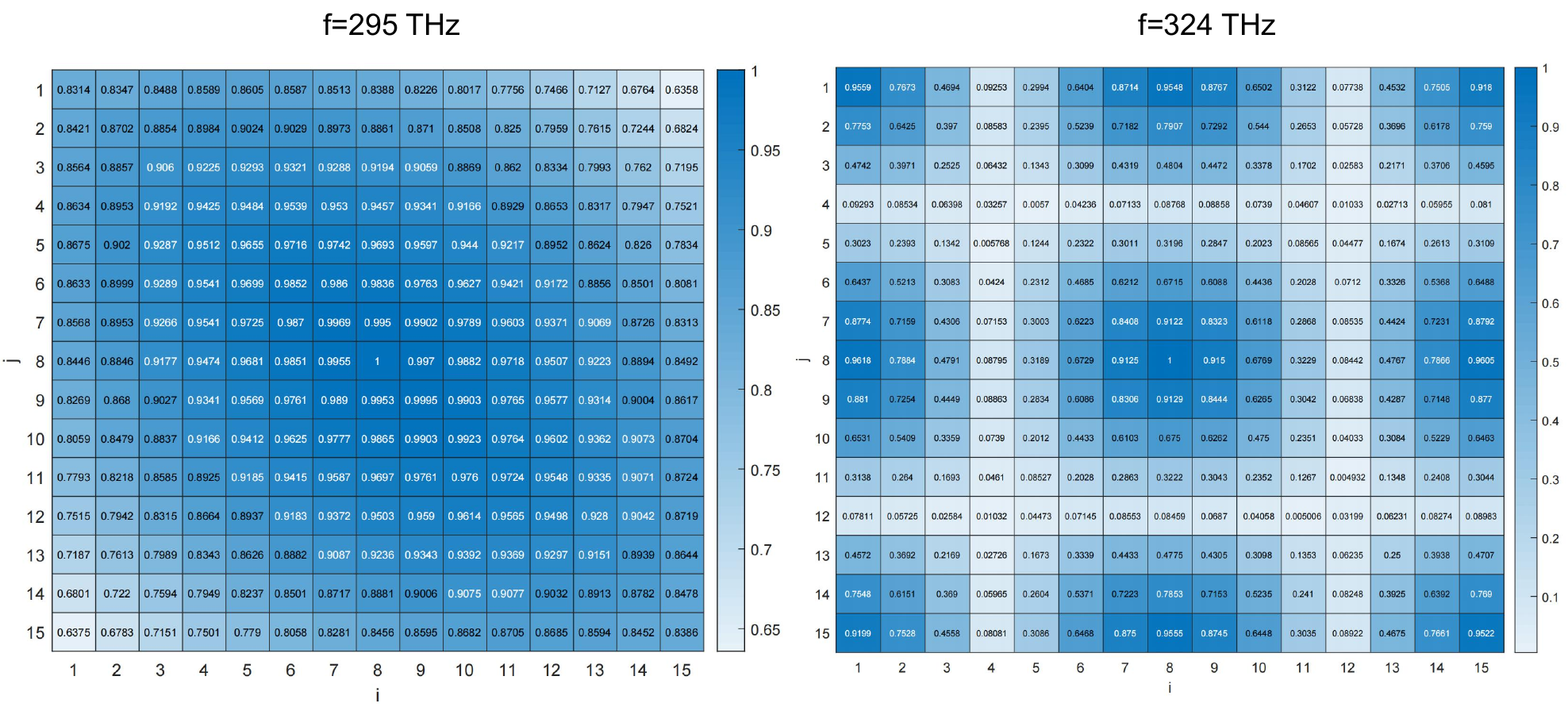,width=0.8\textwidth}
	\caption{(Color online) The calculated decoherence matrix $\boldsymbol \Gamma$ of 15 emitters uniformly located on the central line of the slit waveguide. The transition frequency of the emitters are taken as (left panel) $f=295$~THz that is the cutoff frequency that ENZ occurs, and (right panel) $f=324$~THz which is another resonance frequency but not ENZ.}
\end{figure}

In Fig.~2, we plot the heatmap of the calculated decoherence matrix $\boldsymbol \Gamma$ of 15 emitters uniformly located on the central line of the slit waveguide. The transition frequency of the emitters are taken as (a) $f=295$~THz that is the cutoff frequency that ENZ occurs, and (b) $f=324$~THz which is another resonance frequency but not ENZ. Obviously, in the case of ENZ, the coupling changes gradually with the emitter-emitter distance, while at the non-ENZ resonance frequency, the coupling changes with the distance very drastically.

\subsection*{V. Details of ENZ photonic crystal}

The photonic energy band of the 2D square cylinder photonic crystal is plotted in Fig.~3. The radius of the cylinder $R=85.3$~nm and the refractive index of the cylinder is $n=2.41$, and the surrounding is air. Fig.~3(b) is the calculated refractive index of photonic crystal around the frequency $f=407.1$~THz, as marked with red dotted circles in two panels. The effective index is calculated using the boundary-mode analysis of RF module in COMSOL multiphysics.

\begin{figure}
	\epsfig{figure=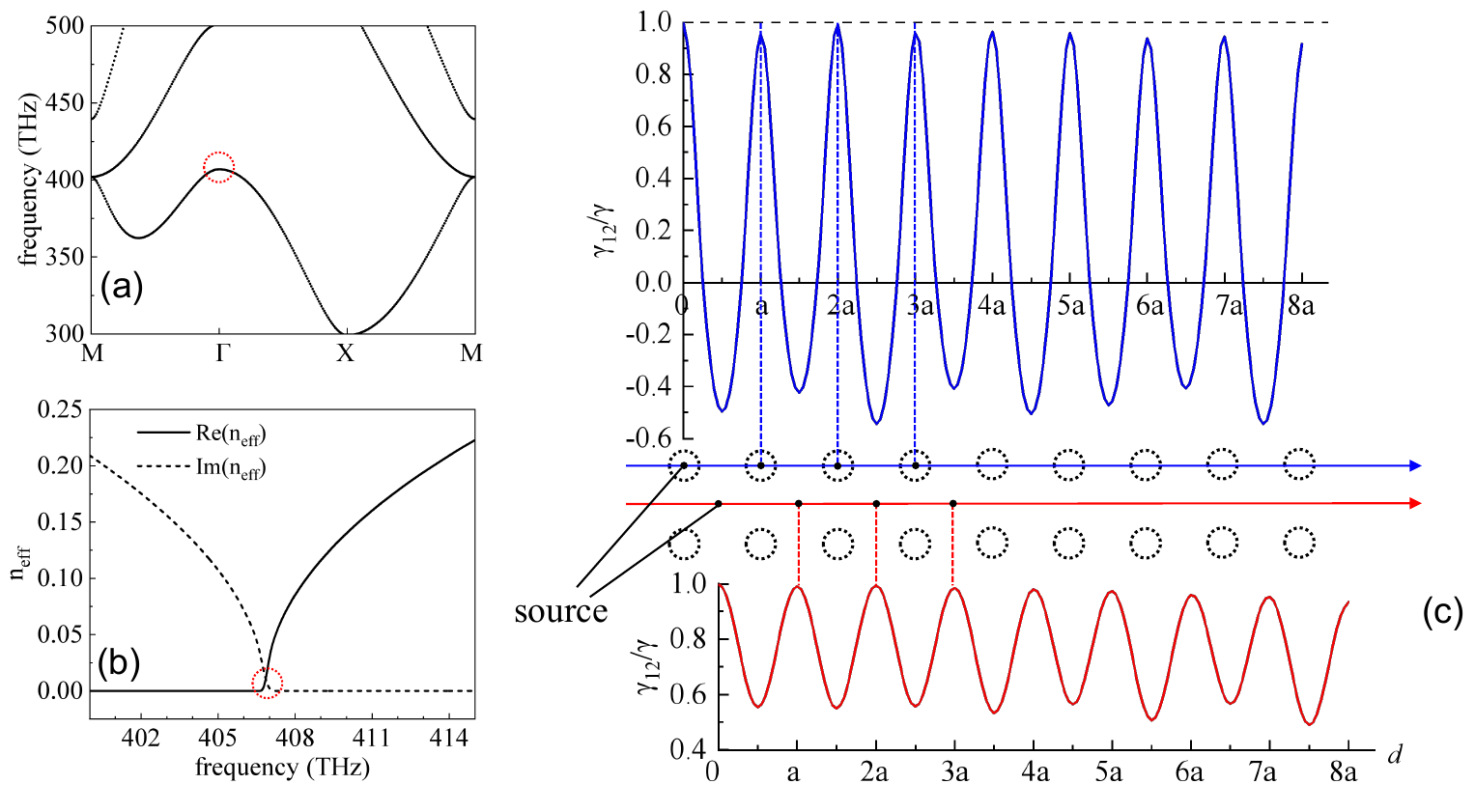,width=0.7\textwidth}
	\caption{(Color online) (a) The photonic energy band of the 2D square cylinder photonic crystal, the lattice constant $a=480.3$~nm. The radius of the cylinder $R=85.3$~nm and the refractive index of the cylinder is $n=2.41$, and the surrounding is air. (b) The calculated refractive index of photonic crystal around the frequency $f=407.1$~THz, as marked with red dotted circles in two panels. (c) The normalized dissipative coupling between two emitters, where an emitter is fixed at the center of a cylinder and the other emitter moves away from the fixed one, see the upper panel plotted with blue line. The case for the line along the cell corner is plotted with red line in the lower panel. }
\end{figure}

In Fig.~3(c), we present the normalized dissipative coupling between two emitters $\gamma_{12}/\gamma_{11}$ as a function of their separation distance $d$ , where the first emitter is fixed at the center of a cylinder (who serves as the source when calculate the field) and the other emitter moves away from the fixed one, see the upper panel plotted with blue line. The lower panel of Fig.~3(c) is the case for the source emitter is fixed at the cell corner, and the other emitter moves along the line of cell corner. Obviously, the coupling of emitter in the case of the upper panel has a larger oscillating with the distance compared with the lower case. Therefore, in order to keep the coupling as uniform as possible, placing the emitter in the air rather than in the cylinder will have less stringent requirements on the position of the emitters.

Figure 4 shows the calculated decoherence matrix $\boldsymbol \Gamma$ of 9 emitters arranged as a 3$\times$3 two-dimensional array, as shown in the inset of Fig.~4(c) of the main text. The left panel of Fig.~4 is the case that emitters located at the center of the cylinders, and the right panel represent the case that emitters are at the corners of the cells.

\begin{figure}
	\epsfig{figure=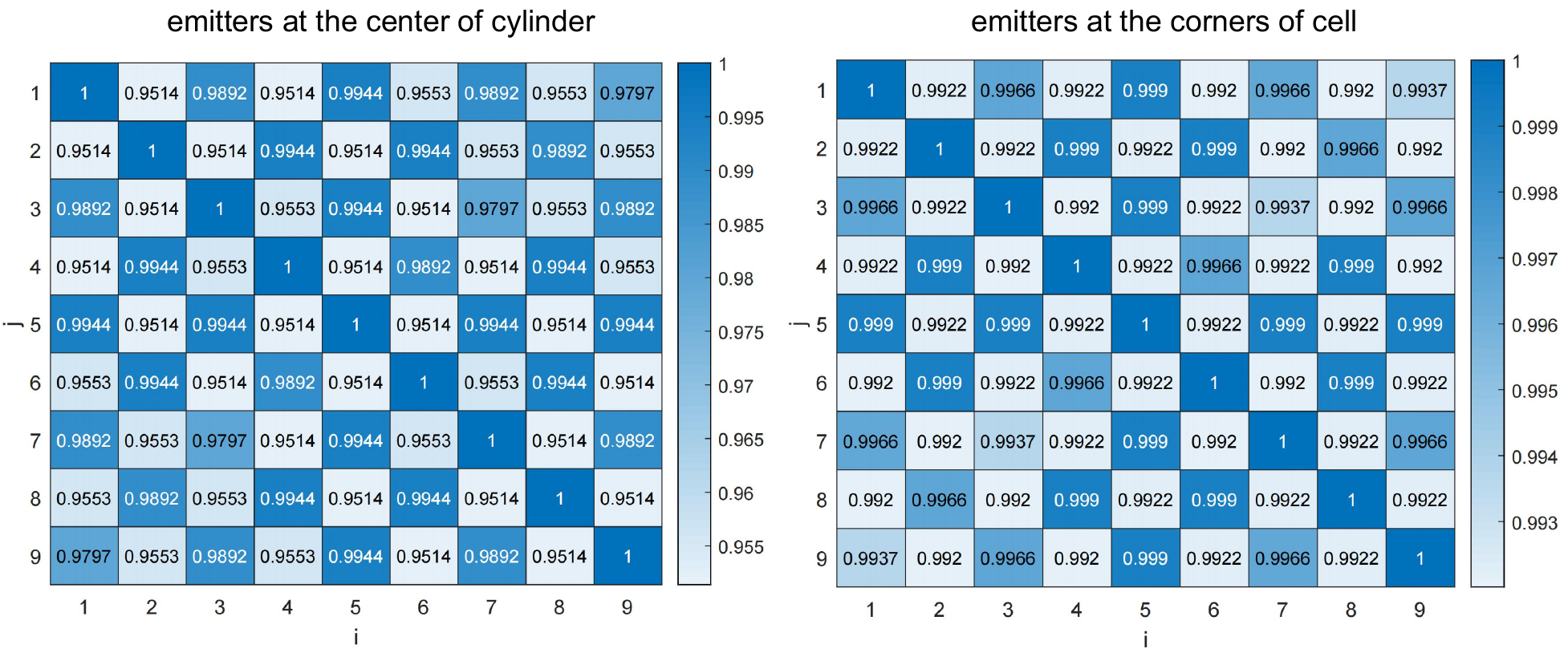,width=0.7\textwidth}
	\caption{(Color online) The calculated decoherence matrix $\boldsymbol \Gamma$ of 9 emitters arranged as a 3$\times$3 two-dimensional array. All the emitters are at (left panel) the center of the cylinder, and (right panel) the corner of the cell.}
\end{figure}

\end{document}